\documentclass[final,3p,times,sort,compress]{elsarticle}

\usepackage{graphicx}
\usepackage{amsmath,amssymb,physics}
\usepackage{hyperref}
\usepackage{xspace}
\usepackage{xcolor}
\usepackage{slashed}
\usepackage{array}
\usepackage{longtable}
\usepackage{subfigure}
\usepackage{pythonhighlight}
\usepackage{ulem}
\usepackage{fontawesome}
\newcommand{\repoURL}{https://github.com/ML4HEP-Theory/ArgoLOOM.git}
\newcommand{\repoText}{\texttt{ML4HEP-Theory/ArgoLOOM}}

\usepackage[T1]{fontenc}
\usepackage{inconsolata}   
\usepackage{fvextra}
\usepackage[most]{tcolorbox}

\DefineVerbatimEnvironment{terminal}{Verbatim}{
  breaklines=true,
  breakanywhere=true,         
  obeytabs=true,
  showspaces=false,
  fontsize=\small,
  tabsize=2,
}

\newtcolorbox{terminalbox}{
  colback=white,
  colframe=black!40,
  boxrule=0.3pt,
  left=2mm, right=2mm, top=1mm, bottom=1mm,
  sharp corners
}

\newcommand{\AL}{\texttt{ArgoLOOM}\xspace}
\newcommand{\CLASS}{\texttt{CLASS}\xspace}
\newcommand{\MG}{\texttt{MadGraph5}\xspace}

\begin{document}

\begin{frontmatter}

\title{\texttt{ArgoLOOM}: agentic AI for fundamental physics from quarks to cosmos}
\tnotetext[t1]{Preprint: ANL-199516}

\author[one]{S.~D.~Bakshi}
\author[two]{P.~Barry}
\author[two]{C.~Bissolotti}
\author[two]{I.~Cloet}
\author[one]{S.~Corrodi}
\author[one]{Z.~Djurcic}
\author[three]{S.~Habib}
\author[one]{K.~Heitmann}
\author[one]{T.~J.~Hobbs}
\ead{tim@anl.gov}
\author[one]{W.~Hopkins}
\author[two]{S.~Joosten}
\author[one]{B.~Kriesten}
\author[three]{N.~Ramachandra}
\author[three]{A.~Wells}
\author[two]{M.~Zurek}

\affiliation[one]{organization={High Energy Physics Division, Argonne National Laboratory},
            city={Lemont},
            postcode={60439}, 
            state={IL},
            country={USA}}

\affiliation[two]{organization={Physics Division, Argonne National Laboratory},
            city={Lemont},
            postcode={60439}, 
            state={IL},
            country={USA}}

\affiliation[three]{organization={Computational Science Division, Argonne National Laboratory},
            city={Lemont},
            postcode={60439}, 
            state={IL},
            country={USA}}

\begin{abstract}
Progress in modern physics has been supported by a steadily expanding corpus of numerical
analyses and computational frameworks, which in turn form the basis for precision calculations and baseline predictions in experimental programs.
These tools play a central role in navigating a complex landscape of theoretical models and current
and potential observables to identify and understand fundamental interactions in physics.
In addition, efforts to search for new fundamental interactions increasingly have a
cross-disciplinary nature, such that understanding and leveraging interoperabilities among computational
tools may be a significant enhancement.
This work presents a new agentic AI framework, which we call~\href{https://github.com/ML4HEP-Theory/ArgoLOOM}{\AL}, designed to bridge methodologies
and computational analyses across cosmology, collider physics, and nuclear science.
We describe the system contours, key internal aspects, and
outline its potential for unifying scientific discovery pipelines.
In the process, we demonstrate the use of~\AL on two small-scale problems to illustrate
its conceptual foundations and potential for extensibility into a steadily growing agentic framework
for fundamental physics.
\end{abstract}


\end{frontmatter}

\section{Introduction}
\label{sec:intro}
The modern quest to interrogate nature and identify novel fundamental physics ranges from femtometer-scale probes of the internal quark-gluon structure of hadrons at collider facilities like the Large Hadron Collider (LHC)~\cite{Feng:2010gw} and the future Electron Ion Collider (EIC)~\cite{AbdulKhalek:2021gbh} to cosmological investigations of the megaparsec-scale correlations in the cosmic microwave
background (CMB) and large-scale structure.
Between these energy extremes lie an array of intersecting domains against
which the standard paradigms of the standard model, $\Lambda_\mathrm{CDM}$, and hadronic/nuclear structure
may be tested using computational physics frameworks of growing complexity.
These computational frameworks have been indispensable in single-domain modeling, {\it e.g.}, setting baselines in
searches for unknown physics or characterizing the behavior of detectors. 
However, the increasing complexity
required to achieve the necessary precision in the quest for New Physics significantly complicates
the optimization of such calculations, signaling a paradigm shift towards multi-domain modeling that exploits the interconnected regions.

In this context, we introduce~\AL ({\it i.e.}, an Argonne-based system of {\bf L}inked {\bf O}racles for
{\bf O}bservables and {\bf M}odels), which is a demonstrator platform for agentic AI for fundamental
physics; in this study, we illustrate~\AL in an intellectual space intersecting currently open problems
in cosmology, particle, and nuclear physics.
At its core, \AL consists of a series of \texttt{Python} steering scripts that interface with the
OpenAI API, using one of the available GPT models as an agentic backbone. This configuration
allows~\AL to initiate a physics dialogue, guide the user in understanding open problems
in fundamental physics, and perform downstream calculations using a suite of computational
modules the agent is configured to invoke and run.
While this initial collection of tools is limited to a minimal set for the sake of demonstrating the
cross-frontier potential of~\AL, the ability of the orchestrator agent to invoke an array of analysis frameworks implies substantial extensibility for future applications.
In addition, while~\AL has been initially developed at Argonne National Laboratory and interacts with Argonne's ``\texttt{Argo}'' system (hence the name~\AL), we release
the initial set of steering codes based on the OpenAI API as suitable for community use.

The remainder of this paper is as follows: after a brief contextual overview of the recent
developments in agentic AI for discovery science in Sec.~\ref{sec:background}, we provide
a high-level summary of the main features of the~\AL framework itself in Sec.~\ref{sec:framework}.
The aspects we discuss include the framework architecture (\ref{sec:arch}), knowledge base (\ref{sec:knowledge}),
and dependence upon chosen backbone model (\ref{sec:backbone}). After this discussion, we
next in Sec.~\ref{sec:modules} present the core computational ingredients within each of the domain-specific physics modules ---
these include cosmology (\ref{sec:cosmo}), particle physics (\ref{sec:collider}), and nuclear physics (\ref{sec:NP}).
Having outlined the~\AL framework itself, we demonstrate its practical use in two complementary
Case Studies in Sec.~\ref{sec:case-studies}: one comparatively top-down illustration (Sec.~\ref{sec:top-down}),
based on a deductive simulation chain assuming a specific, concrete model for BSM physics;
and a more inductive, bottom-up example (Sec.~\ref{sec:bottom-up}), wherein a particular
observable related to nuclear experiments at lower energies is assumed and the potential
model relevance of those observables are logically explored.
In Sec.~\ref{sec:extend}, we address the potential to extend the~\AL framework with, {\it e.g.},
an enlarged basis of modeling tools and reasoning capabilities before concluding
in Sec.~\ref{sec:conclusion}.
%

%
%

\section{Background}
\label{sec:background}
The past $\sim$year has witnessed the rapid emergence of agentic AI
systems applied to an array of scientific tasks including
literature searches, hypothesis formulation and ideation, and the
deployment of complex computational frameworks.
This efflorescence suggests that a significant portion
of physics analysis and technical workflows may be substantially automated.
Agentic AI systems are generally typified by the relative autonomy with which
they can carry out computational tasks with some degree of adaptivity; they
are capable of planning actions and integrating with external libraries, which
might be invoked in accordance with an established reasoning paradigm.
Crucially, their utility extends beyond the invocation of isolated code workflows towards coordinating multi-domain physics modeling.

Agentic AI approaches have been deployed in a range of discovery
science contexts, including the biology of protein design~\cite{ghafarollahi2025sparksmultiagentartificialintelligence},
fundamental chemistry involving the synthesis of novel compounds~\cite{bran2023chemcrowaugmentinglargelanguagemodels,doi:10.1021/acsomega.4c08408},
observatory operations for gamma-ray astronomy~\cite{kostunin2025aiagentsgroundbasedgamma};
similarly, agentic methods have been demonstrated for cosmology~\cite{Laverick:2024fyh,Moss:2025ynt, tam2025infera}, including
the~\texttt{DR MACS} system~\cite{DRMACS}.

\begin{python}
+-------------------------------------------------+
| || || || || || || || || || || || || || || || || |
| || || || || || || || || || || || || || || || || |
     _                    _     ___   ___  __  __ 
    / \   _ __ __ _  ___ | |   / _ \ / _ \|  \/  |
   / _ \ | '__/ _` |/ _ \| |  | | | | | | | |\/| |
  / ___ \| | | (_| | (_) | |__| |_| | |_| | |  | |
 /_/   \_\_|  \__, |\___/|_____\___/ \___/|_|  |_|
              |___/                               

| || || || || || || || || || || || || || || || || |
| || || || || || || || || || || || || || || || || |
+-------------------------------------------------+
|  A r g o L O O M  :  weaving Quarks to Cosmos   |
|  Linked Oracles  for  Observables and Models    |
+-------------------------------------------------+
    
[ArgoLOOM] Ready. Type your request (Ctrl+C to exit).
\end{python}

\section{Agentic AI overview}
\label{sec:framework}
The~\AL package provides a centralized codebase of agentic steering scripts which
allow the iterative invocation of analysis codes for fundamental physics.
Fundamentally, it is a single-agent backbone (by default, \texttt{GPT-4o}) which is invoked
through a driver script which acts as an orchestrator to call a series of physics modules.
In this section, we briefly discuss the configuration of ~\AL in
Sec.~\ref{sec:arch} before drawing attention to the curated knowledge base (\ref{sec:knowledge}) and considerations
related to the choice of backbone model (\ref{sec:backbone}).
%

%
%

\subsection{Framework architecture and tools}
\label{sec:arch}
\begin{figure}[htbp]
  \centering
  \includegraphics[width=0.98\textwidth]{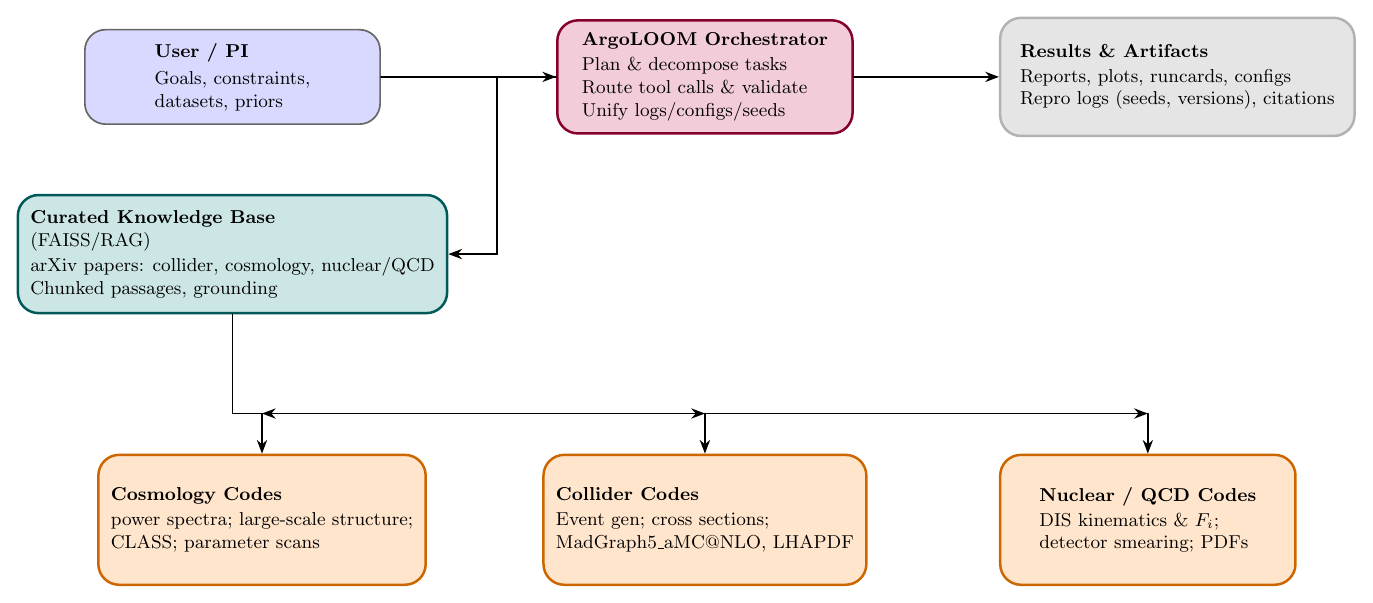}
  \caption{
	  A depiction of the~\AL agentic workflow, focusing on the main components: a primary backbone model to orchestrate calculations and information flow in the cosmology, particle, and nuclear physics domains, as well as interaction with a specialized knowledge base.
	}
  \label{fig:flow}
\end{figure}

As we intend~\AL as an initial demonstration of an agentic AI analysis pipeline for cross-disciplinary
questions in fundamental physics, we choose a fairly simple, parallel structure in which a
small-scale orchestrator rests above domain-specific invocation scripts for
cosmology, nuclear, and particle physics.
The conceptual organization of~\AL is depicted in Fig.~\ref{fig:flow}, which illustrates the main scaffolding of an agentic workflow.
A typical analysis proceeds as follows. 
The user specifies physics inputs such as overarching goals, theory constraints, experimental datasets, and prior distributions. 
These are passed to the~\AL orchestrator, which ingests the input directions and constructs a plan to address the user's task. 
The orchestrator consults the curated knowledge base consisting of arXiv literature in collider phenomenology, cosmology, and QCD. With this knowledge base, the orchestrator consults the user to define the output deliverables the user would like to generate.
It then dispatches workflows through a multi-domain pipeline to produce cosmological power spectra, collider-based new physics signatures, and nuclear/QCD detector-level observables.
Finally, the orchestrator produces reports with citations, plots, runcards, configuration files, and data logs with ready-to-run deliverable scripts to then return the artifacts to the user.
We give a brief overview of the main scripts within the~\AL repository.
\begin{itemize}
  \item \texttt{argo-loom.py} -- the front-end for the pipeline. With this script, the user dialogues with the OpenAI API about model instantiation, specific collider processes, and run parameters.
  \item \texttt{API\_MG5-kin.py} -- the OpenAI API initiates and runs~\MG with a specific UFO model file \cite{Degrande:2011ua}. The script generates an input number of events from the user and parses the cross section. Next it locates the Les Houches files and constructs physics quantities such as rapidity distributions. The script maps these quantities to kinematics like $x_{1,2}$ and $Q^{2}$. 
  \item \texttt{toolkit\_class.py} -- Contains the \CLASS analysis Boltzmann solver to build auto- and cross-spectra by configuring \CLASS runs, deciding on adjustable parameters, and executing the computation. Includes simple plotting utilities to visualize resulting spectra against $\ell$.
  \item \texttt{toolkit\_kb.py} -- Knowledge base lookup utilities; directs the agent toward FAISS chunked indices and queries according to $k$-nearest neighbors ranking metrics; the number of ranked results can be adjusted.
\end{itemize}
In addition to these main steering scripts, the \AL package also contains a number of supporting tools which may be used independently of the dialogue itself; these include builder and querying scripts related to the FAISS knowledge base discussed in Sec.~\ref{sec:knowledge} below; users may leverage these codes to expand their own curated set of references.
%
%

\subsection{Internal knowledge base}
\label{sec:knowledge}

We augment the initial layer of~\AL with a curated knowledge base utility to provide a focused and interpretable
steering mechanism to the agentic workflow.
The library consists of $\sim$5 arXiv papers in each of the three~\AL domains spanning collider
phenomenology, cosmology, and nuclear physics (QCD). We select this specific corpus to ensure robust signal-to-noise
coverage of representative methods and results across these subject areas, rather than an exhaustive breadth. This
approach ensures that retrieved technical passages have direct scientific relevance to the computational workflows
as we describe in more detail in Sec.~\ref{sec:modules}.

Practically, each of the arXiv documents are ingested through a series of \texttt{Python} scripts developed for this
purpose, which in turn perform indexing with Facebook AI similarity search (FAISS) such that texts are segmented into semantically coherent chunks for
efficient vector search (by default, this is performed in an~\AL query tool based on $k$-nearest neighbor lookup). The~\AL
knowledge base is therefore accessed through retrieval-augmented generation (RAG), allowing the agent to identify contextually
relevant passages to inform downstream computations. Crucially, this solution carries the advantage of both reducing the
risk of stochastic hallucination while interpretably providing verifiable citations for reproducibility and transparency.
For the demonstration of~\AL, our knowledge base provides a basis for cross-domain reasoning; for instance,
queries related to sterile neutrinos retrieve constraints or search strategies relevant for both cosmology as well as
high-energy collider phenomenology, while QCD references drawn from nuclear physics literature can inform parton-level
calculations. The modular construction allows straightforward expansion to additional papers, datasets, or
community-specific corpora, ensuring long-term sustainability and extensibility; these latter aspects we
address in more detail in Sec.~\ref{sec:extend}.

%
\subsection{Dependence on backbone model}
\label{sec:backbone}
The improvements afforded by agentic workflows may be connected in part to the
assumed backbone model.
In this study, we explored a range of possibilities, assuming as a default the \texttt{GPT-4o} model.
Relative to this, we also considered a parametrically smaller (and computationally cheaper)
variant, \texttt{GPT-4o-mini}, as well as the recently released \texttt{GPT-5} model, which augments the
reasoning chain capabilities available in \texttt{GPT-4o}.
These variations can be consequential in steering downstream tool invocation.
We note that, while many other alternatives might be considered further --- including
smaller LLMs as well as open-weights models like those of the Llama family --- these three GPT variants bookend a number of scenarios, going
from relatively small-scale models to orchestrate physics tool calls to more recent reasoning
models ({\it e.g.}, \texttt{GPT-5}) capable of informing calculations with a large corpus of physics
logic as well as downstream invocations.
With respect to this latter point as a partial justification, reasoning models have been
suggested~\cite{Saqlain:2025owc} as having discrimination power in New Physics searches
in fairly standard LHC search channels.
Similarly, it has been shown in the context of multi-agent frameworks for collider
analyses that the choice of backbone model can have significant influence on performance
in demonstration tasks like identifying BSM signals in representative search
channels~\cite{Diefenbacher:2025zzn}.
We reserve a more thorough quantitative comparison among these choices to future study, and adopt the \texttt{GPT-4o} model as our default below.


%
\subsection{Getting started: dependencies, configuration, recommendations}
\label{sec:startup}
The pilot release of~\AL is intended as a largely standalone tool which might be run on a local workstation with functioning versions of \MG~\cite{Alwall:2011uj} and \CLASS~\cite{Lesgourgues:2011re} available.
The scripts and interfaces comprising~\AL can be obtained from the dedicated GitHub repository, \href{\repoURL}{\faGithub\ \repoText}, which
also contains a brief README with guidelines on getting started quickly as well as basic usage recommendations.
For stable integrations, we suggest configuring these codes inside a standard \texttt{conda} environment with, {\it e.g.}, \texttt{Python}~\texttt{3.10}.
The scripts which interface the OpenAI API are principally written in \texttt{Python} and require standard libraries and packages which the user should \texttt{pip install}.
These libraries include \texttt{numpy}; \texttt{faiss} (for querying the knowledge base); \texttt{sentence-transformers} and related AI libraries in the \texttt{PyTorch} family; \texttt{pyarrow} for handling of data outputs; \texttt{pylhe} for parsing of Les Houches event (LHE) files; \texttt{matplotlib} (for the automated generation of plots); as well as the \texttt{openai} API library itself.


\section{Domain-specific calculations and modules}
\label{sec:modules}
In this section, we briefly highlight the domain-specific components of the
toolkit as orchestrated by~\AL.
For the initial demonstration of the~\AL approach, relevant actions are commonly used codes or computational methods employed in each domain.
For cosmology, \AL employs the~\CLASS Boltzmann solver as we overview in Sec.~\ref{sec:cosmo}.
The main toolset for HEP collider phenomenology, discussed in Sec.~\ref{sec:collider} is the \MG generator framework, used to compute the partonic hard-scattering interactions relevant for predictions at the LHC.
In Sec.~\ref{sec:NP}, we describe the corresponding computational solvers for nuclear physics in the QCD sector, which consist mainly of kinematic matchings and nuclear science collider detector models required to relate experimental measurements at the high-energy and nuclear frontiers.
Via the agentic orchestrator, these internal modules are capable of interacting, both with each other, as well as with information generated by calls to the knowledge base discussed in Sec.~\ref{sec:knowledge}.
After discussing the internal aspects of these domain-specific kernels and their main assumptions at a high level, we present illustrative case studies in Sec.~\ref{sec:case-studies}.


\subsection{Cosmology Integration}
\label{sec:cosmo}
The initial~\AL integrations for cosmology are substantially related to the
invocation of the \CLASS code for large-scale structure (LSS) in the cosmic microwave
background (CMB).
In particular, \AL furnishes a direct, agentic interface to \CLASS involving the
automated generation of steering card (.ini) inputs allowing the execution of the main release
of the framework, \texttt{class\_public}; in addition, \AL permits quick exploration
and parsing of output spectra as generated by \CLASS.
These spectra are represented through expansions in spherical harmonics of the form
\begin{equation}
    C^{ij}_\ell = {1 \over (2\ell + 1)}\, \sum_m a^i_{\ell m}\, (a^j_{\ell m})^\star\ , \ \ \ \ i, j = \{T, E, B, \phi\}\ ,
\end{equation}
which are associated with temperature ($T$), $E$-mode ($E$), $B$-mode ($B$), and gravitational lensing
($\phi$) maps. As is typical in practice, we plot the $\ell$-weighted spectra of the form
\begin{equation}
    D^{ij}_\ell = \left({1 \over 2 \pi}\right)\ \ell\, [\ell +1] C^{ij}_\ell\ .
\label{eq:D-class}
\end{equation}
Within this basis, a series of auto- and cross-correlations may be evaluated, {\it e.g.}, the
$E$-mode auto spectra, $D^{EE}_\ell$, as well as analogous cross spectra such as the
temperature-$E$-mode quantity, $D^{TE}_\ell$; these are in addition to the auto-spectrum of the
multipole, $\ell$, itself --- $D^L$.
The~\AL utility allows the rapid exploration of a variety of theoretical assumptions available inside
\texttt{CLASS} which can directly impact these spectra, such as the effects of a sterile neutrino as we
discuss in more detail in Sec.~\ref{sec:top-down} below.
Variations in theoretical assumptions can be implemented through adjustments
of the configurable parameters ($N_\mathrm{ncdm}$, $m_\mathrm{ncdm}$, $T_\mathrm{ncdm}$,
$\mathrm{deg}_\mathrm{ncdm}$) to study BSM cosmologies.
In addition, \texttt{ArgoLOOM} contains automated plotting routines allowing the end-to-end production of
CMB angular power spectra ($C_\ell$), with decomposition into TT, EE, TE, and lensing contributions as
well as subsequent quick visualizations.
As with other integrations in the~\AL codebase, these downstream calculations are governed by
interactive steering wherein the backbone agent may be prompted in conversational mode allowing the user to
propose model variants with respect to altered neutrino masses or numbers as well as the use of possible physics
extensions ({\it e.g.}, \texttt{ExoCLASS}~\cite{Stocker:2018avm}) to get fast outputs.
Additionally, the set of codes related to~\CLASS is supplemented by a curated knowledge-base composed of literature
on BSM models and their cosmological implications to guide suggested parameter ranges and their
potential phenomenological relevance.
While this initial release of~\AL interfaces default~\CLASS as a primary demonstration,
it is readily extensible to accommodate additional variants of~\CLASS, such as \texttt{CLASSgal} \cite{DiDio:2013bqa} for
relativistically corrected LSS observables, or CAMB for cross-validation.


\subsection{High-energy collider physics integration}
\label{sec:collider}
AI/ML methods have been developed and used in collider phenomenology since before the deep-learning revolution,
with generative AI specifically playing an important and expanding role~\cite{Almaeen:2024guo, Kriesten:2024ist,Kriesten:2024are,Kriesten:2023uoi,Kriesten:2025gti} in recent years.
Conceptually, the corresponding integrations for the high-energy collider capability in~\AL mirror those outlined briefly
above for cosmology.
In particular, for collider applications --- especially those relevant for the LHC --- \AL functions mainly as a~\MG pipeline orchestrator
wherein the backbone agent generates command scripts, compiles the associated hard-scattering amplitudes for processes specified
in the steering cards, and launches the associated event-generation runs;
from this point, \AL can then direct the location of native~\MG output (LHE files) and perform basic
parsing of this output to explore unintegrated cross sections.
This basic functionality is then augmented by a number of internal routines to facilitate more rapid exploration
of a wide range of possible BSM physics scenarios in the form of a model management utility to direct the
automated fetching and installation of UFO models (via FeynRules \cite{Alloul:2013bka} or custom URLs) into the \MG environment.
The parsing noted above permits fast numerical explorations of the generated events. For instance, \AL can steer the
conversion from LHE outputs to physics observables such as rapidity distributions, invariant masses, and partonic
kinematics internally to inform the interpretation of, {\it e.g.}, possible BSM model dependence or kinematical
regions of more significant sensitivity.
Physics interpretations are then further informed by rapid visualizations of the output generated during runs through
the automated production of histograms, two-dimensional scatter plots, or heatmaps to inspect distributions.
As in the cosmological integrations, these calculations rest within a larger agentic workflow in which chat-driven
adjustments to beam energies, collider processes, BSM parameters can be implemented by the agent
dynamically or multiple physics runs might be chained in dialogue.
Lastly, integration hooks allow results to be exported as, {\it e.g.}, Parquets for further downstream ML training
if so desired.


\begin{figure}[htbp]
  \centering
  \includegraphics[width=0.48\textwidth]{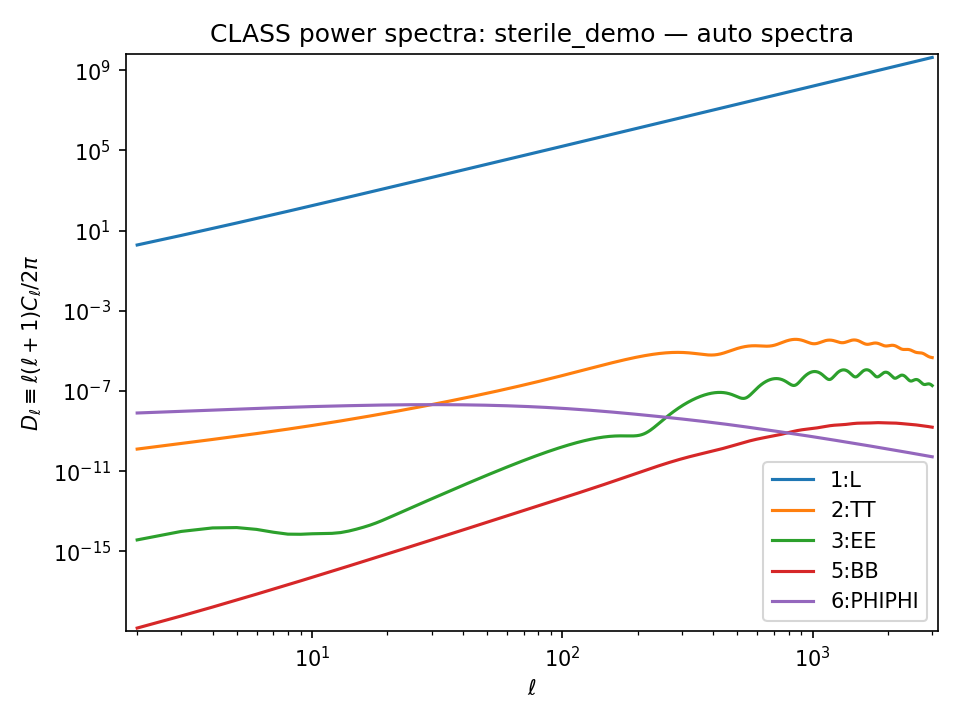}
  \includegraphics[width=0.48\textwidth]{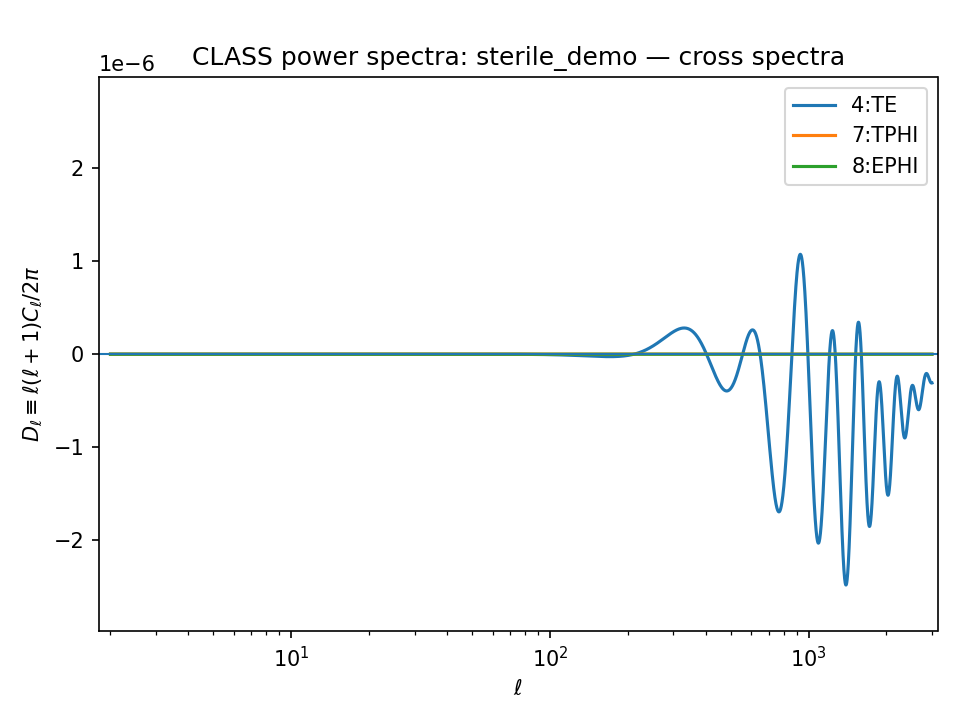}
  \caption{Once~\AL is directed toward a standard~\CLASS build, it can agentically
	setup run cards and executions base on a range of cosmology scenarios
	and microphysics assumptions. These may be steered based on initial
	dialogue with the backbone agent and consultation of the knowledge base.
	In this example, we illustrate a set of power
	spectra (left) generated by~\AL as well as families of deviations
	from these defaults based on sterile neutrino scenarios (right).}
  \label{fig:cosmo-sterile-nu}
\end{figure}

\subsection{Nuclear physics integration}
\label{sec:NP}
The internal nuclear physics (NP) integrations of the~\AL framework resemble those of the
respective cosmology and particle physics (collider components) as discussed above.
Within the NP domain, key computational ingredients include calculations related to deeply inelastic scattering
(DIS) from the proton. Associated with these are routines for interpreting and
parsing LHE output from~\MG, and then extracting from these kinematical information relevant for QCD quantities
like PDFs or other hadron-structure correlation functions.
In the latter case, \AL can map simulated events typical of high-energy colliders like the LHC to parton-level kinematics
relevant to DIS experiments like those at the EIC~\cite{AbdulKhalek:2021gbh}.
At these matched kinematics, \AL can also invoke basic smearing models to estimate the resolution effects in DIS detectors
which complicate the extraction of parton-level information, either in the standard model ({\it e.g.}, the PDFs) or possible
BSM signatures.
In this study, these kinematical matchings are to the partonic momentum fractions of the respective colliding protons at the
LHC ($x_1$ and $x_2$) as well as to the DIS virtuality, $Q^2$.
With these capabilities, \AL can kinematically map at Born-level, {\it e.g.}, the rapidity ($y$)
distributions as generated within~\MG in its HEP collider domain to partonic variables to which NP DIS
experiments are sensitive as at the EIC. For example, for $pp \to Z$ production at Born-level:
\begin{equation}
    x_{1,2} = {M_Z \over \sqrt{s}}\ \exp\, (\pm y)\ ;\ \ \ \ Q = M_Z\ ,
\end{equation}
where $\sqrt{s}$ indicates the center-of-mass energy of the $pp$ collision --- for instance, 13 TeV
at the LHC.
These matching calculations are thus carried out in a routine
which connect LHC observables like rapidity into NP-relevant parton fractions, $x_{1,2}$ and virtualities
$Q^2$ over which DIS measurements might take data with modeled detector smearing.
Such matching calculations are in addition to the hard-scatter modeling of DIS itself as available in~\MG, which can
be agentically carried out at EIC or other NP kinematics, already providing some ability to simulate
electron-ion collisions or generate pseudodata for impact studies.
We note that these DIS calculations also permit direct calculations over a wide array of available BSM
theories and parametrizations (including SMEFT) and permit the simultaneous exploration of New Physics scenarios,
including some which might be motivated by independent information from cosmology. As we address regarding
extensibility in Sec.~\ref{sec:extend}, it is possible to broaden this base of New Physics assumptions going forward,
for instance, by including an expanded set of higher-dimension operator insertions for DIS processes.
Similarly, both the direct hard-scatter modeling and kinematical matching and detector tools permit various
possible PDF sensitivity studies; for instance, it is possible to generate automated comparisons of DIS/EIC theory
replicas (on the NP side) or LHC predictions against multiple PDF sets --- a feature which is useful for
exploring SM uncertainties related to hadronic structure.
As with the cosmology and HEP integrations, the nuclear physics elements of~\AL are supplemented by
the knowledge base discussed in Sec.~\ref{sec:knowledge}, which can provide valuable steering with respect to
DIS detector knowledge or interplay with QCD information and connections to legacy ({\it e.g.}, BCDMS, HERA, or SLAC)
experiments via curated ingestion into the knowledge base.
%


\section{Case Studies}
\label{sec:case-studies}
In this section we present two representative illustrations of~\AL, meant to highlight the conceptual possibilities and cross-frontier applications of an agentic AI workflow as well as technical aspects of the associated physics calculations discussed in the previous sections.
The first, described in Sec.~\ref{sec:top-down}, amounts to a problem scenario in which a specific BSM theory is known ({\it e.g.}, in terms of its Lagrangian) and the user would like to explore the top-down phenomenological consequences, such as computing numerical predictions for different measured quantities in collider experiments.
The second is agnostic with respect to any particular underlying B/SM mechanism. In this scenario, described in Sec.~\ref{sec:bottom-up}, a dialogue may start from an empirical measurement at energy regimes well below the TeV-scale. The user can then use ~\AL to explore possible SM backgrounds for the experimental observables in question as well as potential BSM scenarios to which they might be sensitive.
The iterative, or ``thinking", mechanism of agentic AI, manifest in~\AL, is such that the domain specialist (the user) can explore new physics scenarios in both directions (top-down or bottom-up) at any time. 
The two case studies demonstrated below may differ in their starting point and motivations; however, exploratory cycles are possible and even inevitable in which, for instance, an initially top-down study based on an assumed BSM scenario might implicate particular observables, which could be interrogated further, leading to entirely different BSM theory computations or SM background projections from the bottom-up.
These dynamic shifts in analysis paradigm are enabled through the agentic dialogue the user initiates with~\AL and represent the strength and flexibility of such a research methodology.

\begin{figure}[tbp]
  \centering
  \includegraphics[width=0.48\textwidth]{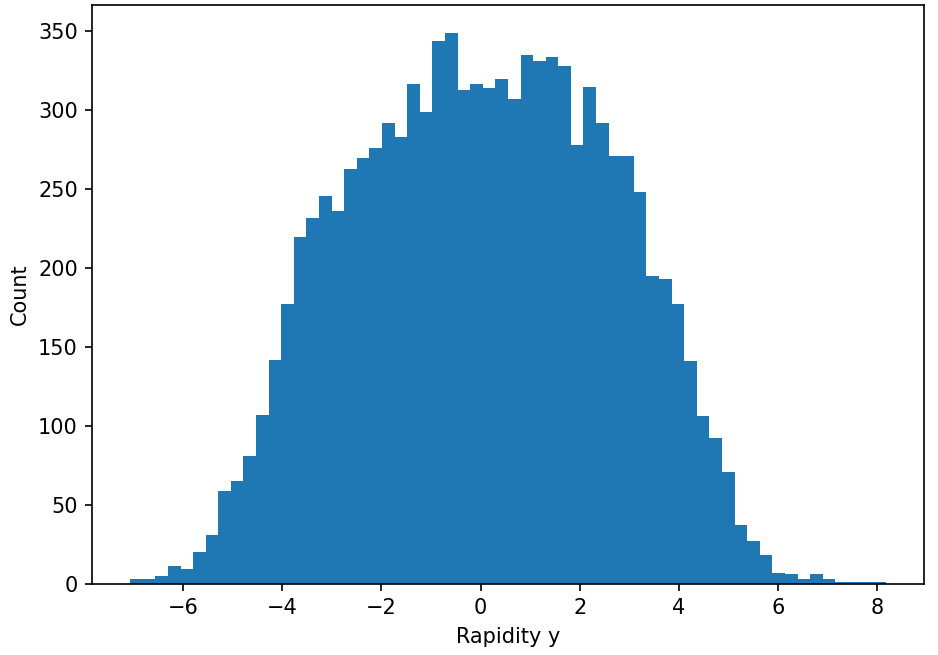}
  \includegraphics[width=0.48\textwidth]{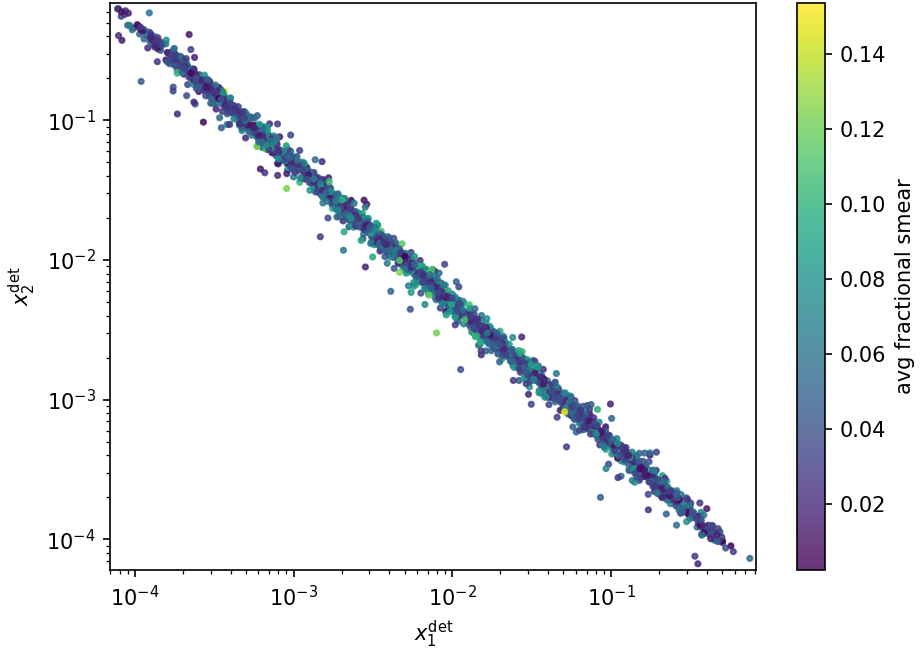}
  \caption{
	  Mirroring the \CLASS setup, \AL can also invoke and steer computations
	  relevant for particle-nuclear physics. For instance, these can include the
	  assumption of a definite SM or BSM theory baseline ({\it i.e.}, corresponding
	  to a particular Lagrangian), from which collider simulations and kinematical
	  matchings to observables relevant to nuclear science might be performed. We
	  illustrate this with a rapidity distribution in $y_{\ell^+\ell^-}$ for 13 TeV
      neutral-current Drell-Yan (left)
	  and corresponding kinematical matchings relevant for DIS observables with
      modeled detector-smearing effects (right), both produced
	  in automated fashion within~\AL.
	}
  \label{fig:part-nucl}
\end{figure}


\subsection{Case 1: BSM theory to cross-frontier observables}
\label{sec:top-down}
As a first demonstration, we deploy \texttt{ArgoLOOM} in a top-down fashion by assuming a specific underlying physics scenario ({\it i.e.}, at the Lagrangian-level) and initiate an~\AL dialogue.
As depicted in Fig.~\ref{fig:flow}, the exploration chain begins with a call to the backbone model. 
The orchestrator then queries the internal knowledge base and launches the calculation sequence using the simulation toolkit discussed in Sec.~\ref{sec:modules} to evaluate the implications for actual measured observables.
We choose sterile neutrinos~\cite{Boyarsky:2018tvu} as a compelling, specific New Physics scenario with cross-domain
relevance to both cosmology~\cite{Boyarsky:2009ix} as well as particle-nuclear physics~\cite{Deppisch:2015qwa}.
Sterile neutrinos have attracted sustained interest given their potential role as a minimal extension of the standard
model. They modify the electroweak sector via
\begin{align}
\mathcal{L} &= \mathcal{L}_{\rm SM}\
  +\ \frac{1}{2}\,\overline{N}_i\,i\slashed{\partial}\,N_i\,
  -\, \bigg(y_{\alpha i}\,\overline{L}_\alpha\,\widetilde H\,N_i
      + \frac{1}{2}\,\overline{N}_i^{\,c}\,(M_N)_{ij}\,N_j + \text{h.c.}
    \bigg)\, .
\label{eq:SN-lagrange}
\end{align}
Through the additional mass and interaction terms, sterile neutrinos can account for the mass hierarchy among active neutrinos via the seesaw mechanism;
contribute to generating the baryon asymmetry; and act as dark-sector portals over a wide
range of mass scales, running from $\mathrm{keV}$ to $\mathrm{TeV}$. Such features may
leave detectable signatures in terrestrial experiments --- for instance, in the form of displaced vertices in
heavy neutral lepton searches at the LHC --- and induce large-scale structure effects in
the cosmic microwave background as relevant to early universe cosmology.
The phenomenological breadth of sterile neutrinos distinguishes them as an ideal candidate to demonstrate an agentic AI
analysis platform. These hypothetical particles can be sought across multiple experimental contexts and possible mass windows of relevance. Consequently, analysis pipelines spanning cosmology, high-energy colliders, and nuclear-QCD physics can be coordinated in a campaign of related calculations to explore the implications of Eq.~\ref{eq:SN-lagrange}.
This BSM case study is one of hundreds which~\AL can explore and extend through open-ended analysis dialogues.
Studies with~\AL are adaptable to varying levels of physics expertise by combining guidance
from the \texttt{GPT} backbone and a curated knowledge base with direct access to computational tools. An expert
user might already maintain a collection of BSM scenarios and associated model (UFO) files in place with corresponding parameter
settings, and thus use~\AL to automate and expedite a series of simulation runs.
Alternatively, a user less acquainted with the subtleties of specific BSM model building may initiate an investigation from a more
general set of questions or assumptions, refining these through interaction with the agent before initiating specific computations.
To illustrate these points, we show the start of the corresponding~\AL dialogue in truncated form in the first set of
terminal output of this section.
%

\begin{terminalbox}
\begin{terminal}
[ArgoLOOM] Ready. Type your request (Ctrl+C to exit).

> I would like to explore the implications of a possible sterile neutrino for cosmology, collider phenomenology at the LHC, and related kinematics at a DIS experiment. Can you help me perform some relevant simulations after consulting the knowledge base?

[agent] tool call $\to$ kb\_search {"index\_dir":"kb\_out","query":"sterile neutrino implications in cosmology and collider phenomenology"}

[assistant]: Sterile neutrinos, which do not interact via the standard weak interactions, have significant implications in both cosmology and particle physics, particularly at colliders and deep inelastic scattering (DIS) experiments. Here's an exploration based on current literature:

1. **Cosmology**:  
In cosmology, sterile neutrinos are a candidate for dark matter. They could have been produced in the early universe through various mechanisms, such as resonance production, non-resonant production, or decay of a heavier particle [doi: 10.1016/j.ppnp.2018.10.002](https://arxiv.org/abs/1807.07938).

They can influence the formation of structures in the universe and can be constrained by observations of cosmic microwave background anisotropies and large-scale structures [doi: 10.1016/j.physletb.2005.11.014](https://arxiv.org/abs/astro-ph/0503065).

2. **Collider Phenomenology**:  
At colliders like the LHC, sterile neutrinos can be searched for through processes that result in lepton number violation or through missing energy signatures associated with their production and decay. Heavy sterile neutrinos could provide observable effects in precision measurements and through rare decay processes [doi: 10.1016/j.plb.2004.02.036](https://arxiv.org/abs/hep-ph/0602150).

3. **DIS Experimentation**:  
In the context of DIS experiments, sterile neutrinos can alter the expected neutrino scattering rates and spectral features due to their potential mixing with active neutrinos. These changes can be probed through precision measurements of neutrino-nucleus interactions.

Let's proceed with simulations for these scenarios. For cosmology, we'll use CLASS to simulate the CMB power spectrum with sterile neutrinos. For collider phenomenology and DIS, we will employ MadGraph to explore heavy sterile neutrino signatures. Please provide the necessary paths for executing CLASS and MadGraph, and let me know if you have specific parameters in mind or would prefer defaults.
\end{terminal}
\end{terminalbox}
%
%

As can be seen in the dialogue shown above, prompting~\AL with a very general query regarding the foundations and testability of sterile neutrinos leads to a check of the internal knowledge base.
Of particular notability is the relevant references to arXiv preprints from which the orchestrator pulls. These references may be consulted manually by the user, or through subsequent~\AL prompts for a more detailed description of interesting BSM scenarios, excluded regions of the parameter space, or current exploratory strategies --- including those of a cross-frontier nature.
As can be seen in the latter portion of the~\AL output above, the system recognizes the request to perform a numerical simulation. In this case the orchestrator calls~\CLASS and~\MG in parallel to simulate CMB power spectra or collider observables, respectively.
From this point, \AL may be prompted further to set up either one or a series of these numerical simulations.

In the case of the cosmology simulations with~\CLASS, \AL can potentially start a power spectrum calculation using assumed defaults for input parameters or may ask the user to specify these. In addition, the current version of~\AL prompts the user to identify the local build location of~\CLASS itself.
In the case of the sterile neutrino scenario, \AL identifies and primarily shifts the value of $N_\mathrm{eff}$ as the relevant parameter. Doing so results in power-spectrum predictions like those
shown in Fig.~\ref{fig:cosmo-sterile-nu}. This figure shows typical~\CLASS spectra for the quantities of Eq.~\ref{eq:D-class} --- giving the $i\!=\!j$ auto-spectra in the left panel and the $i\!\neq\!j$ cross spectra in the right panel.

Meanwhile, starting from the same fundamental sterile neutrino hypothesis, \AL can additionally
launch simulations to explore sensitivity at colliders.
Mirroring the \CLASS setup, \AL prompts the user to identify a local installation of~\MG before inquiring about other inputs of the collider simulation such as collision energies, $\sqrt{s}$, event numbers, $N_\mathrm{event}$, and the collider process itself ({\it e.g.}, Drell-Yan dilepton production, $t\bar{t}$ processes, or DIS). \AL also requires knowledge of the underlying theoretical assumptions regarding the specific BSM scenarios, perturbative order, theory input parameters, parton distribution functions (PDFs) and other relevant settings.
In this latter case, \AL checks the \texttt{model/} directory of~\MG against the desired process specified by the user. In the example above, which already assumes a sterile neutrino hypothesis, \AL can consult the \texttt{model/} directory for the specific model files implicated by the foregoing sterile neutrino discussion, if available.
Alternatively, the user may also wish to perform a background study of purely standard model processes in the signal region. The~\AL pipeline can configure such runs assuming the \texttt{sm/} UFO files.
We illustrate the result of this particular case in Fig.~\ref{fig:part-nucl} (left), wherein we plot the rapidity distribution, $y_\ell$, of the final-state lepton produced in charge-current Drell-Yan, $pp \to Z \to \ell^+\, \ell^-$, on the logic that the tails of such distributions may be sensitive to $\sim$GeV-scale sterile neutrino masses and beyond.

Finally, nuclear physics integrations in the QCD sector allow the results of HEP collider
simulations to be related to observables of relevance to experiments at lower-energy kinematics, like DIS measurement campaigns at the planned EIC.
For instance, Fig.~\ref{fig:part-nucl} (right) illustrates a Born-level kinematical
matching of a rapidity distribution like that shown in the left-hand panel to corresponding
values of parton fractions, $x_1$ and $x_2$, of the colliding protons; these are then
mapped into a two-dimensional space showing how the parton-level kinematics probed at high-energy
colliders project to regions of phase space which might be explored at the future EIC. It is also possible to compute detector-level effects in this space according to simple smearing models as shown here.

We reiterate that this set of calculations in each of cosmology, particle, and nuclear physics are emblematic of an iterative chain of potential calculations and exploratory investigations which would continue from this point. For instance, having identified a region of the $x_{1,2}$ space implicated by the sterile neutrino simulation, one can then ask~\AL whether other BSM scenarios may be similarly sensitive to this region and/or whether there may be cosmological signatures associated with these models. Alternatively, one might further investigate the SM backgrounds through an extended series of additional~\MG simulations, and ensemble these results for further inspection.
As such, an initially top-down problem framing as discussed here might naturally lead to a bottom-up series of inquiries as we discuss in the following Sec.~\ref{sec:bottom-up}.


\subsection{Case 2: Medium-energy observables to TeV-scale models}
\label{sec:bottom-up}
To complement the top-down demonstration of Sec.~\ref{sec:top-down}, we also consider
a scenario in which~\AL starts from a different sector of the cosmology-nuclear-particle
landscape --- that is, a problem that does not assume a specific fundamental interaction
at high energies ({\it i.e.}, a top-down approach), but starts from a phenomenological
observable at lower energy in order to investigate the high-energy scenarios to which
this might be sensitive in parallel with collider and cosmology information.

In this scenario, a more bottom-up agentic dialogue may begin, not from an assumed high-energy interaction
and corresponding UFO file in \MG, but from an empirical observable associated with, {\it
e.g.}, a QCD-sensitive process at a nuclear physics experiment such as EIC-like DIS
collisions.
In this respect, such a study would effectively begin from the latter stage(s) of the workflow in the more top-down case study discussed in Sec.~\ref{sec:top-down}.
A typical user in this scenario might, for instance, be a researcher in a field adjacent to HEP collider physics interested in exploring to what extent measurements or theoretical calculations in their home domain ({\it e.g.}, nuclear [QCD] physics) might involve crosstalk with collider phenomenology or cosmology.
In such a case, an~\AL study might start from a general question about the sensitivity DIS measurements, characteristic of the EIC, might hold in terms of BSM scenarios or related backgrounds.
The out-of-the-box response of~\AL following a prompt of this nature is shown below.

\begin{terminalbox}
\begin{terminal}
	[ArgoLOOM] Ready. Type your request (Ctrl+C to exit).

> I would like to perform a more ground-up study to explore what possible BSM signatures might be implicated by the kinematical region of high-x for the PDFs and a scale of 10 GeV. Can you help guide me?
[agent] tool call --> kb_search {"index_dir":"kb_out","query":"high-x PDFs BSM signatures"}

[assistant]: To investigate possible Beyond Standard Model (BSM) signatures at high momentum fractions (high-x) with a scale of 10 GeV, it's important to consider how high-x phenomenology could be related to BSM physics, especially in processes like Drell-Yan where rapidity or invariant mass distributions are involved at large scales. According to recent literature, experiments like those at the Electron-Ion Collider (EIC) aim to reduce limitations on Parton Distribution Functions (PDFs) by improving sensitivity to these high-x regions, potentially revealing new physics signatures [EIC White Paper, arXiv:2203.13199].

You might want to run simulations using a tool like MadGraph to explore specific processes at a scale of 10 GeV while focusing on higher momentum fractions to see if unexpected events appear, which could hint at BSM physics. Let me know if you'd like to proceed with simulations using these tools!

> Great! It seems that high-x at large scales in Drell-Yan might be relevant. Do you have any suggestions for the rapidity to explore?
[agent] tool call --> kb\_search {"index\_dir":"kb\_out","query":"Drell-Yan high-x rapidity"}

[assistant]: In Drell-Yan processes, especially when exploring BSM physics at high-x and large scales, the choice of rapidity plays a significant role in the sensitivity to possible new signals. The rapidity or invariant-mass distributions in these processes at large mass scales are crucial areas of study [EIC White Paper, arXiv:2203.13199].

If you're considering a simulation, selecting rapidity values on the order of high absolute values (e.g., around +/- 2 to +/- 3, assuming the experimental setup permits) might yield interesting results. Please let me know if you'd like to proceed with a simulation based on these inputs! 

> Thank you --- perhaps I might go with SUSY in this case. Are there any SUSY scenarios which might impact LSS in the context of CLASS calculations?
[agent] tool call --> kb\_search {"index\_dir":"kb\_out","query":"SUSY dark matter LSS CLASS"}

[assistant]: Supersymmetry (SUSY) offers several intriguing scenarios for dark matter candidates, which could have implications on large-scale structure (LSS) formation and be modeled using tools like CLASS.

These scenarios impact LSS by altering the dark matter distribution, which in turn affects the growth of cosmic structures. When running CLASS simulations, adjusting parameters like the density of dark matter and its interaction cross-sections would be necessary to explore these effects.

If you wish to run a CLASS simulation considering one of these SUSY scenarios, let me know how you'd like to proceed with the parameters!
\end{terminal}
\end{terminalbox}

In this interaction with~\AL, combined access to the reasoning abilities of the \texttt{GPT} backbone model and curated knowledge base allows the user to ascend an inductive chain. In practice, this entails flowing from a general observable (DIS structure functions at $Q\!\sim\!10$ GeV and high $x$), to high-energy collider processes (Drell-Yan $pp$ events), to kinematical regions of cross sections for these processes, such as $d\sigma /dy$ for $|y|\!\sim\!2\!-\!3$.
Moreover, from this stage it also becomes possible to prompt~\AL further regarding specific BSM scenarios --- with the framework in this case pointing to several supersymmetric models as possibilities for simulation.
%


\section{Extensibility}
\label{sec:extend}
This first release of~\AL is a pilot demonstration of a cross-frontier agentic AI
for fundamental physics.
It has been constrained to a relatively small code and knowledge base which
is nevertheless targeted to illustrate a proof-of-concept.
For this reason, opportunities exist to extend~\AL in several topical
directions as well with respect to core capabilities.

The reasoning capacity in~\AL is significantly augmented
by the physics knowledge base which partly steers the downstream invocation of
computational tools. For this initial study, the knowledge base was highly
focused on problems directly related to, {\it e.g.}, sterile neutrino scenarios
at colliders and in cosmology, as well as crossover issues at the interface of
DIS (nuclear physics) and high-energy experiments.
This knowledge base might therefore be grown systematically to encompass a wider
range of problems related, {\it e.g.}, to testing the standard model in a large
setting of cross-frontier experiments and observables.

With respect to the internal toolkit available to~\AL, there are various opportunities for expansion.
In many physics cases, additional capabilities might be deployed within the same basic architecture embodied by Fig.~\ref{fig:flow}, adding further modules for specialized tasks which might be pulled by the orchestrator in the same manner as the dedicated scripts for~\CLASS, \MG, and the nuclear/QCD kinematical maps and related tools.
Short-term expansions might involve folding in extensions of the core~\CLASS framework like the~\texttt{ExoCLASS} plug-in for handling additional non-standard physics beyond $\Lambda_\mathrm{CDM}$.
With respect to collider simulations, the minimal~\MG implementation might be extended to more fully automate the sampling of BSM scenarios or generation of predictions based on fast samplings of higher-dimensional SMEFT operator combinations available in, {\it e.g.}, \texttt{SMEFT@NLO} \cite{Degrande:2020evl}.
The nuclear physics and QCD components of~\AL might similarly be broadened by including additional routines to steer PDF selections and uncertainty quantification through looping calculations over \texttt{LHAPDF} error sets and computing additional correlations.
Similarly, the extension to other nonperturbative QCD distributions, such as transverse- or generalized parton distributions, and their relation to collider observables or other cross sections of cosmological relevance might be implemented and investigated.
Widening the scope of calculations which~\AL may perform agentically will permit a considerably expanded set of hypotheses with potential cross-frontier sensitivity to be explored.
Moreover, deploying~\AL on HPC architectures will facilitate the preparation of an extended series of run campaigns, for instance, to explore cross-domain inference with a suite of uncertainties.


\section{Conclusion and Outlook}
\label{sec:conclusion}
Efforts to test modern fundamental physics --- from the standard model
of particle physics, to the $\Lambda_\mathrm{CDM}$ paradigm of cosmology,
to our picture of the internal structure of hadrons and nuclei --- will
only continue to grow in scope and complexity in coming years.
Exploring this richer landscape will necessitate the intelligent and
reproducible deployment of computational and knowledge-based approaches,
the use of which can be significantly aided by agentic AI methods.
In this study, we have introduced the~\AL approach as an end-to-end, cross-frontier
agent with the ability to interweave calculations and reasoning tasks spanning
cosmology and particle-nuclear physics.
In this context, \AL represents an early agentic approach
to computation in fundamental physics, emphasizing high-level planning
and execution across core parts of the scientific pipeline, beyond
simple task automation.
Although at an early stage, \AL incorporates physics-aware ingredients
in the form of RAG-style queries over an internal knowledge base to
inform reasoning and theoretical predictions, while preserving elements
of reproducibility like run cards, Monte Carlo seeds, and specific
citations to arXiv documents that trace its workflow.

These aspects of~\AL offer several advantages, particularly as the framework is extended to include added capabilities as discussed in Sec.~\ref{sec:extend}. For example, an AI agent can automate and  explore BSM scenarios and their implications across cross-frontier experiments and observatories, while enabling uncertainty quantification through iterative physics runs.
In addition, deploying model predictions at scale can facilitate the production of significant volumes of training data for physics-based AI models, including foundation models.
Uniquely, the~\AL cross-frontier toolkit supports extensive agentic dialogues to explore physics hypotheses intersecting multiple frontiers and generates predictions whose correlated signatures can be examined across energy scales or measurement contexts.

Looking ahead, we expect subsequent iterations of~\AL and related agentic AI methods to play a valuable role in cross-frontier studies of fundamental physics.
The established codebase supporting~\AL enables validation of theoretical predictions produced within agentic workflows; such validation is useful for benchmarking purposes and will also be important in ensuring the `AI safety' of agentic approaches in both physics as well as related scientific domains.


\section*{Acknowledgments}
This work was supported by the Argonne Laboratory Directed Research and Development
(LDRD) project, AI for Fundamental Physics: Quarks to Cosmos. We thank Carlos Wagner for helpful discussions on possible cross-frontier sensitivities to sterile neutrino scenarios. Work at Argonne National Laboratory was supported by the U.S. Department of Energy, Office of High Energy Physics. Argonne, a U.S. Department of Energy Office of Science Laboratory, is operated by UChicago Argonne LLC under contract no. DE-AC02-06CH11357.


\bibliographystyle{elsarticle-num}
\bibliography{Q2C}
\end{document}